\documentclass[aps,
scriptaddress,spanish,nofootinbib,prd,preprintnumbers,showkeys,superscriptaddress,longbibliography
Sorrit] 
{revtex4-2}  
\usepackage{amsmath, amssymb}
\usepackage{makeidx}
\usepackage{amsfonts}
\usepackage{ucs}
\usepackage[utf8]{inputenc}
\usepackage[usenames,dvipsnames]{pstricks}
\usepackage{subfigure}  
\usepackage{epsfig}
\usepackage{pst-grad} 
\usepackage{pst-plot} 
\usepackage[colorlinks,hyperindex]{hyperref}
\usepackage{multirow}   
\hypersetup
{
	colorlinks,%
	citecolor=black,%
	linkcolor=black,%
	urlcolor=black,%
}

\usepackage[mathscr]{euscript}
\usepackage{amsmath}
\usepackage{graphicx}
\usepackage{dcolumn}
\usepackage{bm}
\usepackage{epsfig}
\usepackage{amssymb,latexsym,mathrsfs}
\usepackage{graphicx}
\usepackage{color}
\usepackage{hyperref}
\usepackage{float}
\usepackage{diagbox}
\hypersetup{
    colorlinks=true,
    linkcolor=red,
    citecolor=blue,
} 

\usepackage{orcidlink}
\usepackage{comment}

\usepackage{amsmath}
\usepackage{amssymb}
\usepackage{subfigure}
\usepackage{hyperref}
\usepackage{url}
\usepackage{xcolor}
\usepackage{color}
\definecolor{amaranth}{rgb}{0.9, 0.17, 0.31}
\definecolor{purple(munsell)}{rgb}{0.62, 0.0, 0.77}
\definecolor{americanrose}{rgb}{1.0, 0.01, 0.24}
\definecolor{palatinateblue}{rgb}{0.15, 0.23, 0.89}
\definecolor{royalblue(web)}{rgb}{0.25, 0.41, 0.88}
\definecolor{hanpurple}{rgb}{0.32, 0.09, 0.98}
\definecolor{beaublue}{rgb}{0.74, 0.83, 0.9}
\definecolor{carminered}{rgb}{1.0, 0.0, 0.22}
\definecolor{brightpink}{rgb}{1.0, 0.0, 0.5}
\definecolor{vividviolet}{rgb}{0.62, 0.0, 1.0}
\hypersetup{ linktoc=all,
    colorlinks, linkcolor={palatinateblue},
    citecolor={brightpink}, urlcolor={amaranth}}

\makeindex

\begin{document}

	\title{\large Frenet-Serret equations with variable proper acceleration in Minkowski spacetime}
\newcommand{\IPICYTaffil}{Instituto Potosino de Investigación Científica y Tecnológica (IPICYT),\\
San Luis Potosí 78216, México.}

\author{Iv\'an P\'erez-Rom\'an\, \orcidlink{0009-0008-8517-0848}}
\affiliation{\IPICYTaffil}

\author{Michael R.R. Good\, \orcidlink{0000-0002-0460-1941}}
\email{muon@asu.edu}
\affiliation{Physics Department \& Energetic Cosmos Laboratory, Nazarbayev University,\\
Astana 010000, Qazaqstan.}
\affiliation{Leung Center for Cosmology and Particle Astrophysics, National Taiwan University,\\
Taipei 10617, Taiwan.}
\affiliation{Beyond Center for Fundamental Concepts in Science, Arizona State University,\\
Tempe AZ 85287, USA.}

\author{Yen Chin Ong\, \orcidlink{/0000-0002-3944-1693}}
\email{ongyenchin@nuaa.edu.cn}
\affiliation{Center for the Cross-disciplinary Research of Space Science and Quantum-technologies (CROSS-Q),\\College of Physics, Nanjing University of Aeronautics and Astronautics,\\
Nanjing 211106, China.}

\author{Haret C. Rosu\, \orcidlink{0000-0001-5909-1945}}
\email{hcr@ipicyt.edu.mx}
\affiliation{\IPICYTaffil}
\affiliation{Leung Center for Cosmology and Particle Astrophysics, National Taiwan University,\\
Taipei 10617, Taiwan.}


	\begin{abstract}
We study Frenet-Serret equations for timelike worldlines in Minkowski spacetime with proper-time-dependent curvature and torsion. This corresponds to relativistic motion with non-uniform proper acceleration and, when torsion is included, to trajectories whose Frenet-Serret frame rotates beyond the acceleration plane. Using the Gram-Schmidt construction of the tetrad from the four-velocity and its derivatives, we relate the intrinsic Frenet-Serret parameters to kinematic quantities such as proper acceleration, four-jerk, and four-snap. We then consider simple analytic cases for the jerk invariant and torsion, obtaining explicit curvature profiles and reduced Frenet-Serret equations. These examples clarify how non-constant acceleration and torsion modify the geometry of accelerated relativistic motion.
\end{abstract}
\date{\today}
\maketitle
\section{Introduction}\label{sec:intro}

The motion of accelerated observers in relativistic spacetime is most naturally described in terms of quantities defined with respect to proper time. Among these, the four-velocity, four-acceleration, and higher proper-time derivatives, such as the four-jerk, provide a kinematic description of the worldline. A complementary and geometrically intrinsic description is provided by the Frenet-Serret (FS) frame, whose scalar invariants characterize the local shape of the worldline independently of a particular coordinate parametrization. In this language, the curvature is identified with the magnitude of the proper acceleration, while the torsion and hypertorsion describe how the FS frame rotates beyond the plane generated by the four-velocity and four-acceleration.

The FS frame was originally introduced for curves in three-dimensional Euclidean space, where curvature and torsion characterize the bending and twisting of a spatial curve. Its generalization to relativistic spacetime provides a useful invariant framework for studying timelike worldlines. In four-dimensional Minkowski spacetime, the corresponding FS system contains three intrinsic parameters: the curvature $\kappa(s)$, torsion $\tau(s)$, and hypertorsion $\nu(s)$, all functions of the proper time $s$ in the most general case. When these quantities are constant, the resulting worldlines are stationary in the sense relevant to relativistic detector theory and quantum field theory in non-inertial frames.

A systematic use of the FS formalism in this context was developed by Letaw \cite{letaw1981stationary}, who classified stationary worldlines in Minkowski spacetime and related them to the vacuum excitation of non-inertial detectors. Related FS constructions have also been considered in curved spacetimes, for example in Ref.~\cite{paithankar2019bound,Shahzad:2024wrg}. The special case of uniform proper acceleration corresponds to the familiar hyperbolic, or Rindler, trajectory \cite{rindler1960hyperbolic,rindler1966kruskal,semay2006observer,perezrrosu2022}. More general stationary worldlines, with constant intrinsic FS invariants, have been studied in connection with vacuum fluctuations, detector response, and Unruh-like effects \cite{doi:10.1063/1.525364,Rosu:1999ad,rosu2005quantum,good2019stationary,Good:2020hav}.

By contrast, worldlines with non-constant FS invariants are less commonly treated in explicit analytic form. They are of interest in the present work. Physically, allowing the curvature to vary corresponds to allowing a time-dependent proper acceleration. Allowing the torsion to vary introduces additional structure, since the motion is no longer confined to the two-dimensional plane determined by the four-velocity and four-acceleration. In such cases, the higher kinematic quantities, especially the four-jerk and four-snap, are not merely formal derivatives but encode invariant information about the evolution of the FS frame.

Here, we study FS equations in Minkowski spacetime with non-constant intrinsic parameters, focusing on the relation between the FS invariants and the usual kinematic quantities of relativistic motion. We construct the orthonormal tetrad by applying the Gram-Schmidt procedure to the four-velocity, four-acceleration, and four-jerk. This construction makes explicit the relation between the curvature and the proper acceleration, and it also gives invariant relations involving the jerk modulus, the curvature, the torsion, and their proper-time derivatives.

The main goal is to identify analytically tractable cases in which the curvature and torsion can be obtained from simple assumptions about the jerk invariant. We first consider the curvature-only case, where the jerk invariant satisfies
\[
||J||^2=\kappa^4-\dot{\kappa}^2,
\]
where $||J||^2$ is the magnitude of the four-vector of the jerk, defined as $||J||^2=J_\nu J^\nu$. We then extend the analysis to include torsion, for which the corresponding relation becomes
\[
||J||^2=\kappa^4-\dot{\kappa}^2-\kappa^2\tau^2.
\]
These equations allow one to proceed in two complementary ways: either prescribing the acceleration and computing the jerk invariant, or imposing a simple form of the jerk invariant and solving for the curvature. We focus in particular on the cases $||J||^2=0$ and constant nonzero $||J||^2$, together with simple choices of the torsion. We work with several related kinematic four-vectors. The four-velocity is expressed as $U^\mu$; the four-acceleration is the proper-time derivative of the four-velocity, $A^\mu=\dot{U}^\mu$; similarly, the four-jerk is the derivative of the four-acceleration, $J^\mu=\dot{A}^\mu$; and the four-snap is the derivative of the four-jerk, $S^\mu=\dot{J}^\mu$.

The paper is organized as follows. In Section~\ref{sec2}, we introduce the generalized FS equations in Minkowski spacetime and establish the notation. In Section~\ref{sec3}, we consider the curvature-only case, construct the corresponding tetrad, derive the relation between curvature and proper acceleration, and analyze the cases of vanishing and constant nonzero jerk invariant. In Section~\ref{sec4}, we include torsion and study how the jerk invariant and FS equations are modified. Several explicit cases are considered, including constant torsion and torsion proportional to curvature. Section~\ref{sec5} summarizes the results and discusses their physical interpretation.

\section{Frenet-Serret equations with non-constant parameters}\label{sec2}

The Frenet-Serret (FS) frame provides an intrinsic description of a worldline in terms of an orthonormal basis that is transported along the curve. In four-dimensional Minkowski spacetime, the FS equations for a timelike worldline may be written as \cite{letaw1981stationary}
\begin{align}
\dot{\Lambda}_0^\mu &= \kappa(s)\Lambda_1^\mu, \nonumber\\
\dot{\Lambda}_1^\mu &= \kappa(s)\Lambda_0^\mu+\tau(s)\Lambda_2^\mu, \nonumber\\
\dot{\Lambda}_2^\mu &= -\tau(s)\Lambda_1^\mu+\nu(s)\Lambda_3^\mu, \nonumber\\
\dot{\Lambda}_3^\mu &= -\nu(s)\Lambda_2^\mu .
\label{ec1}
\end{align}
Here \(s\) is the proper time, a dot denotes differentiation with respect to \(s\), and \(\{\Lambda_a^\mu\}\) is the FS tetrad. The three scalar functions \(\kappa(s)\), \(\tau(s)\), and \(\nu(s)\) are the curvature, torsion, and hypertorsion, respectively. They are intrinsic invariants of the worldline. The curvature measures the magnitude of the bending of the trajectory in spacetime, while the torsion and hypertorsion describe successive rotations of the FS frame away from the lower-dimensional planes generated by the proper-time derivatives of the worldline.

We use the metric signature \((+,-,-,-)\). Greek indices label spacetime components, while Latin indices label elements of the tetrad. The tetrad is orthonormal,
\begin{equation}\label{ec2.5}
\Lambda_{a\mu}\Lambda_b^\mu=\eta_{ab},
\end{equation}
with
\[
\eta_{ab}={\rm diag}(1,-1,-1,-1).
\]
The first tetrad vector is chosen to be the four-velocity,
\begin{equation}
\Lambda_0^\mu=U^\mu,
\end{equation}
so that \(U_\mu U^\mu=1\). If all FS invariants vanish, the four-velocity is constant and the worldline is inertial,
\begin{equation}
x^\mu(s)=x_0^\mu+U_0^\mu s .
\end{equation}

Equivalently, Eq.~(\ref{ec1}) can be written in compact form as
\begin{equation}
\dot{\Lambda}_a^\mu=K_a{}^b\Lambda_b^\mu .
\end{equation}
It is often useful to display the antisymmetric lower-index matrix \(K_{ab}\), defined by \(K_a{}^b=K_{ac}\eta^{cb}\). With the sign convention used above,
\begin{align}
\label{ec2}
K_{ab}=
\begin{pmatrix}
           0 & -\kappa(s) & 0 & 0 \\
           \kappa(s) & 0 & -\tau(s) & 0 \\
           0 & \tau(s) & 0 & -\nu(s) \\
           0 & 0 & \nu(s) & 0
         \end{pmatrix}.
\end{align}
This form makes explicit the Lorentz-antisymmetry of the FS transport matrix,
\begin{equation}
K_{ab}=-K_{ba},
\end{equation}
which guarantees that the orthonormality condition in Eq.~(\ref{ec2.5}) is preserved along the worldline.

In the following sections, we construct the tetrad by applying the Gram-Schmidt procedure to the four-velocity and its proper-time derivatives. This makes the connection between the intrinsic FS invariants and the kinematic quantities of relativistic motion explicit. In particular, the curvature will be identified with the proper acceleration. We first consider the case in which only the curvature is present, and then extend the analysis to include torsion. The hypertorsion is displayed here for completeness but is set to zero in the explicit cases studied below.

\section{Curvature}\label{sec3}

The simplest non-trivial case of the FS equations, where only curvature is present, implies a matrix $K$ of size $2\times 2$
\begin{align}\label{ec3}
K_{ab}=\begin{pmatrix}
           0 & -\kappa(s)\\
           \kappa(s) & 0
         \end{pmatrix}~.
\end{align}

In this case, there are only two nonzero elements of the tetrad we need to construct. This also means that there are only two FS equations, $a=0$ and $a=1$ in Eq.~(\ref{ec1}). We consider the four-velocity as the first element
\begin{equation}\label{ec4}
\Lambda_0^\mu=U^\mu~,
\end{equation}
and we construct the next member of the basis as
\begin{equation}
(\Lambda_1^\mu)_{nn}=\dot{\Lambda}_0^\mu-\frac{
\dot{\Lambda}_0^\nu\Lambda_{0\nu}}
{\Lambda_{0\nu}\Lambda_0^\nu}\Lambda_0^\mu,
\end{equation}
where the sub-index $nn$ refers to not being normalized, meaning that the orthonormal condition Eq.~(\ref{ec2.5}) is not yet fulfilled, and the dot refers to the time derivative with respect to the proper time $s$. Taking the derivative of $U_\mu U^\mu=1$ we see that $U_\mu A^\mu=0$. Then, we have that $(\Lambda_1^\mu)_{nn}=A^\mu-U_\mu A^\mu=A^\mu$, and considering the normalization condition Eq.~(\ref{ec2.5}), we obtain the normalized component~
\begin{equation}\label{ec16}
\Lambda_1^\mu=\frac{A^\mu}{\sqrt{-A_\nu A^\nu}}~.
\end{equation}
The acceleration modulus is $A_\mu A^\mu=-\alpha^2$, with $\alpha=\alpha(s)$ being the proper acceleration. Then, we write $\Lambda_1^\mu=A^\mu/\alpha$ and the quantity we need
to replace in Eq.~(\ref{ec1}) is its derivative
\begin{equation}\label{ec17}
\dot{\Lambda}_1^\mu=\frac{d}{ds}\left(\frac{A^\mu}{\alpha}\right)=\frac{J^\mu}{\alpha}-\frac{\dot{\alpha}}{\alpha^2}A^\mu=\frac{J^\mu}{\sqrt{-A_\nu A^\nu}}+\frac{(A_\nu J^\nu)A^\mu}{(-A_\nu A^\nu)^{3/2}}~,
\end{equation}
where we obtain $A_\nu J^\nu=-\alpha\dot{\alpha}$ from $U_\nu A^\nu=0$. Using the first FS equation,
\begin{equation}\label{ec18}
\dot{\Lambda}_0^\mu=\dot U^\mu=A^\mu
=\kappa\Lambda_1^\mu
=\kappa\frac{A^\mu}{\alpha}~,
\end{equation}
so that
\begin{equation}
\kappa(s)=\alpha(s).
\end{equation}
Thus, the curvature equals the proper acceleration, even when the acceleration is not constant.

Using the second FS equation together with Eq.~(\ref{ec17}), one obtains
\begin{equation}\label{ec19}
\frac{J^\mu}{\kappa}
-\frac{\dot{\kappa}}{\kappa^2}A^\mu
=
\kappa U^\mu~,
\end{equation}
which can be written as
\begin{equation}\label{ec110}
J^\mu-\frac{\dot{\kappa}}{\kappa}A^\mu-\kappa^2U^\mu=0~.
\end{equation}
In the latter equation, if $\alpha=\kappa=$ constant, we recover the corresponding FS equation of constant proper acceleration.\\
The standard parametrization $U^\mu \to (\cosh f,\sinh f)$ \cite{perezrrosu2022,ponspalol2019}, where $f(s)$ is known as the rapidity, helps to split Eq.~(\ref{ec110}) into a scalar time component and a spatial component. For the time component, Eq.~(\ref{ec110}) gives
\begin{equation}\begin{split}
&\left(\ddot{f}-\frac{\dot{\kappa}\dot{f}}{\kappa}\right)\sinh f+(\dot{f}^2-\kappa^2)\cosh f=0~,
\end{split}\end{equation}
which vanishes if $\dot{f}(s)=\alpha(s)$. This means that the parametrization is valid if
\begin{equation}\label{fa12}
f(s)=\int\kappa(s)ds+c_2~.
\end{equation}
The spatial components behave similarly, and we obtain the same expression as Eq.~(\ref{fa12}). 

We can compute the jerk modulus from Eq.~(\ref{ec110})
\begin{equation}\label{jec13}
J_\nu J^\nu=||J||^2=\kappa^4-\dot{\kappa}^2~.
\end{equation}
Note that we now have a relation between acceleration, its derivative, and jerk. We can work in two ways: if we know the explicit form of the acceleration, we can easily find the jerk modulus; if we assume a specific form for the jerk, we obtain a differential equation for the acceleration. We consider the cases of a null four-vector $||J||^2=0$ and a constant $||J||^2$.

\subsection{Null vector case: $||J||^2=0$}

\begin{figure}[h]
\includegraphics[scale=0.48]{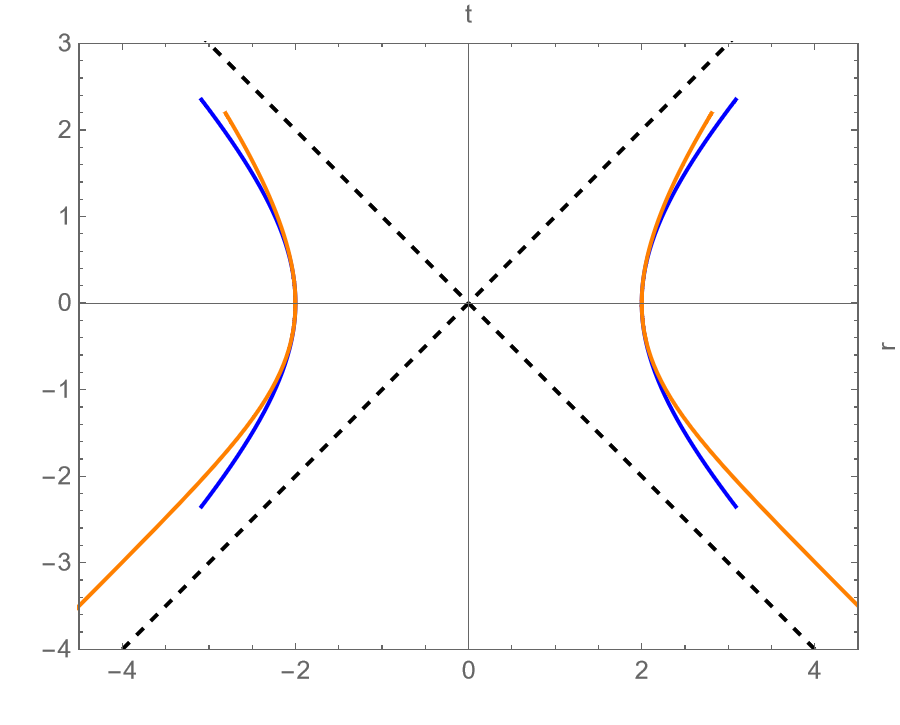}
\includegraphics[scale=0.39]{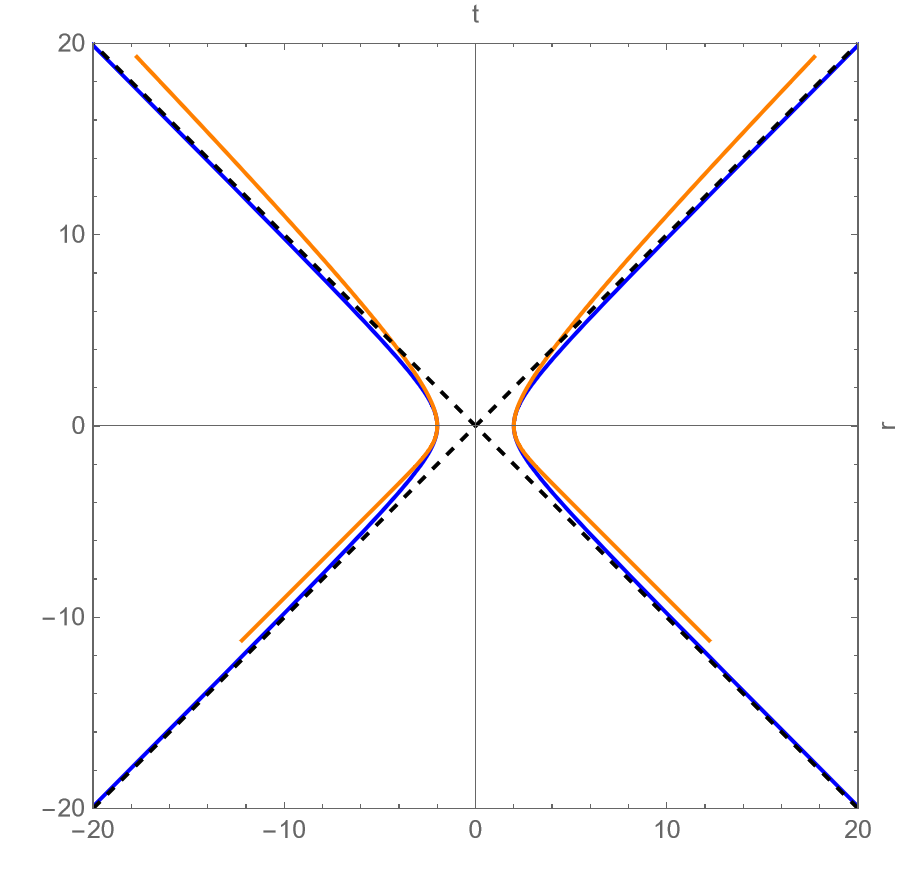}
\caption{Parametric plots of the non-uniformly accelerated trajectories in Eq. (\ref{trzero}), with $\kappa_0=0.5$, compared with the uniformly accelerated Rindler hyperbolas. The left panel uses $s\in(-3,3)$, while the right panel uses $s\in(-30,30)$. The blue curves are the Rindler trajectories
$t_R(s)=\sinh(\alpha s)/\alpha$ and $r_{R,\pm}(s)=\pm\cosh(\alpha s)/\alpha$, with $\alpha=0.5$. The orange curves are the two branches of Eq. (\ref{trzero}), shifted only in the spatial coordinate so that their initial points coincide with the corresponding Rindler vertices: $\tilde r_+(s)=r_+(s)+2$ for the right branch and $\tilde r_-(s)=r_-(s)-2$ for the left branch. Here $2=1/\kappa_0=1/\alpha$, for visual comparison.}
\label{paramscm}
\end{figure}

For this particular case, and considering the result from the first FS equation $\alpha(s)=\kappa(s)$, we have that
\begin{equation}
||J||^2=\kappa^4-\dot{\kappa}^2=0~,
\end{equation}
which provides $\kappa(s)$ by solving $\int\kappa^{-2}d\kappa=\mp\int ds$,
\begin{equation}\label{Eq15}
\kappa(s)=\pm\frac{1}{s+c_1}~,
\end{equation}
where $c_1$ is an integration constant determined by the initial condition. For instance, the condition $\kappa(s=0)=\pm\kappa_0$, with $\kappa_0>0$, gives $c_1=1/\kappa_0$. The particular case where $c_1=0$ implies $\kappa(s=0)\to\infty$, which is the Carlitz-Willey trajectory for a moving mirror which is eternally thermal \cite{carlitz1987reflections}, see also \cite{Good_2018}. Thus, with $c_1=1/\kappa_0$, Eq. (\ref{Eq15}) becomes
\begin{equation}\label{Eq16}
\kappa(s)=\pm\frac{\kappa_0}{1+\kappa_0 s}~,
\end{equation}
there is no singularity for positive proper time, $s\geq 0$. We compute $f(s)$ through Eq.~(\ref{fa12})
\begin{equation}\label{fsint}
f(s)=\int\kappa(s)ds+c_2=\pm\ln(1+\kappa_0 s)+c_2~,
\end{equation}
where the initial condition $f(0)=0$ gives $c_2=0$. Eq.~(\ref{fsint}) implies that one can write the two-velocity as
\begin{align}\label{2vel}
U(s)=\begin{pmatrix}
           \dot{t}(s) \\
           \dot{r}(s)
         \end{pmatrix}=
\begin{pmatrix}
           \cosh(f(s)) \\
	   \sinh(f(s))
         \end{pmatrix}=
\begin{pmatrix}
           \frac{1+(1+\kappa_0s)^2}{2(1+\kappa_0s)} \\
	    \\
           \frac{\mp1\pm(1+\kappa_0s)^2}{2(1+\kappa_0s)}
         \end{pmatrix}~.
\end{align}
Integrating the temporal and spatial components of the velocity, we obtain $t(s)$ and $r(s)$
\begin{equation}\label{trzero}
t(s)=\frac{1}{4\kappa_0}\left((\kappa_0s)^2+2\kappa_0s+2\ln(1+\kappa_0s)\right)~, \ \ \ {\rm and} \ \ \ r(s)=\pm\frac{1}{4\kappa_0}\left((\kappa_0s)^2+2\kappa_0s-2\ln(1+\kappa_0s)\right)~,
\end{equation}
where we have set the integration constants to $-1/4\kappa_0$ and zero, respectively, which matches what is usually done for constant proper acceleration in hyperbolic relativistic motion.
The plots of Eq. (\ref{trzero}) are shown in Figure~(\ref{paramscm}) for two ranges of $s$. Note that in the first case, the behavior is similar to that of a Rindler hyperbola, but it changes in the extended range, where a crossing of the horizon occurs. This is due to the opposite effects of the logarithmic term in the two components of the trajectory, arising from non-uniform acceleration in Eq. (\ref{Eq16}).

\subsection{Case: $||J||^2$ non-zero constant}

\begin{figure}[h]
\includegraphics[scale=0.5]{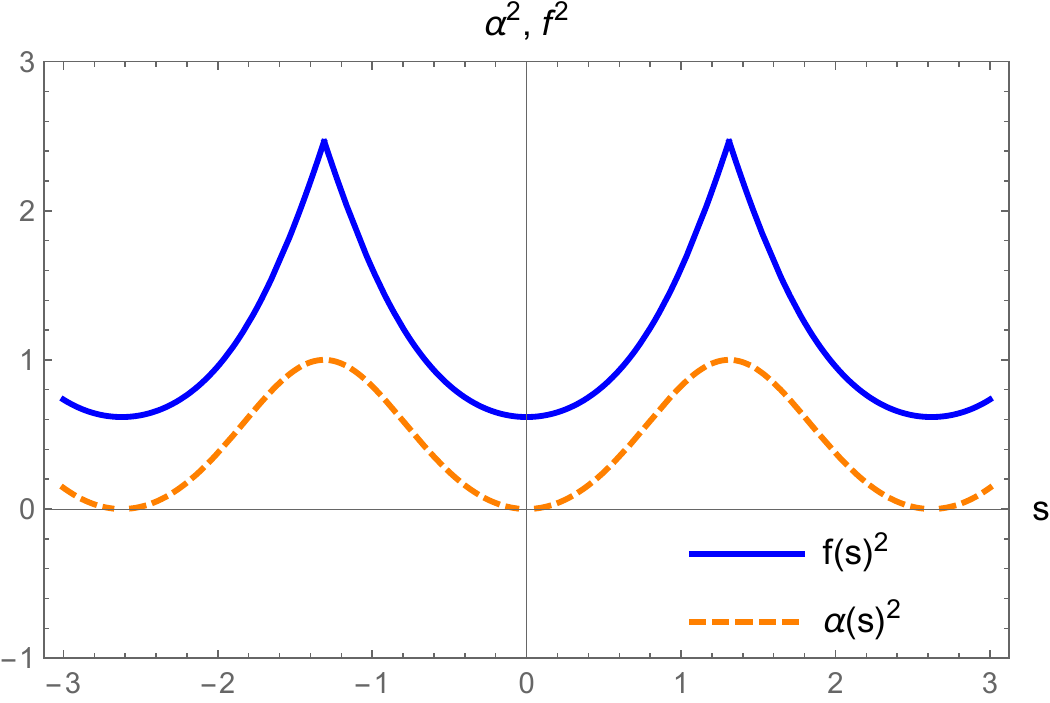}
\includegraphics[scale=0.5]{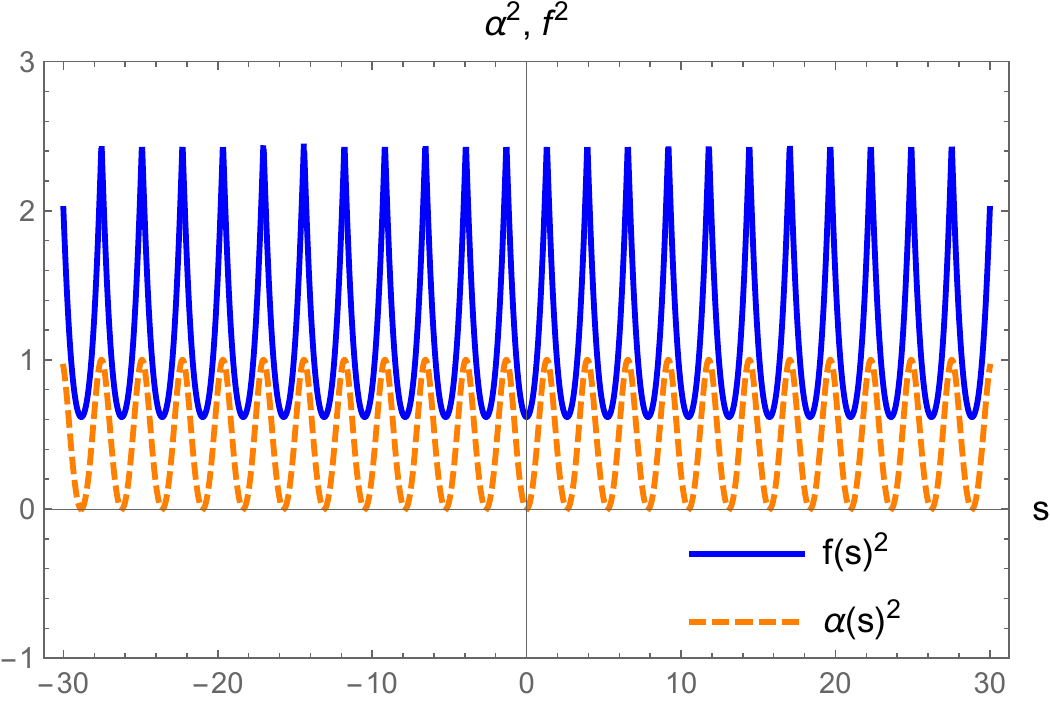}
\caption{Plots of $|\alpha|^2$ and $|f|^2$ in $s\in(-3,3)$ and $s\in(-30,30)$ with $J=1$ and $c_1 = 0$. The periodicity of proper time for both proper acceleration and rapidity underscores the characteristic behavior of the Jacobi elliptic functions under a constant nonzero jerk.}
\label{sn1}
\end{figure}

Writing the jerk modulus as $||J||^2=\kappa^4-\dot{\kappa}^2$ and integrating, one obtains
\begin{equation}\label{j2ncint}
\int\frac{d\kappa}{\sqrt{\kappa^4-||J||^2}}=\pm s+c_1,
\end{equation}
where $c_1$ is an arbitrary integration constant 
that can be fixed by initial conditions. The left side of Eq.~(\ref{j2ncint}) is an elliptic integral of the first kind \cite{zbMATH03863589},
which allows us to obtain $\alpha$ in terms of $s$ as a Jacobi elliptic sine function,
\begin{equation}
\alpha(s)=\kappa(s)=\mp\sqrt{||J||}\,{\rm sn}(i\sqrt{||J||}(s+c_1),-1)~,
\label{paJnonzero}\end{equation}
where $||J||=\sqrt{||J||^2}$. This particular solution represents a complex analytic branch of the elliptic solution, for which the proper acceleration is purely imaginary for real \(s\). By using Eq.~(\ref{fa12}), one can find $f(s)$ in terms of the Jacobi elliptic function
\begin{equation}
f(s)=\mp i\,{\rm arctan}\!\left({\rm cd}(i\sqrt{||J||}(s+c_1),-1)\right)+c_2~,
\label{rapidityJnonzero}
\end{equation}
where $f(s)$ is the rapidity, the argument of the hyperbolic functions that parametrize the velocity four-vector. One must integrate those functions to find the trajectory, which can only be done numerically.  See Figure (\ref{sn1}) for an illustration of the modulus squared of the proper acceleration, Eq.~(\ref{paJnonzero}), and rapidity Eq.~(\ref{rapidityJnonzero}).

\section{Curvature and torsion}\label{sec4}

In the presence of torsion, the matrix $K$ is of size $3\times 3$
\begin{align}\label{ec121}
K_{ab}=\begin{pmatrix}
           0 & -\kappa(s) & 0\\
           \kappa(s) & 0 & -\tau(s)\\
           0 & \tau(s) & 0
         \end{pmatrix}~.
\end{align}
The first two components of the tetrad are already provided by Eq. (\ref{ec4}) and Eq. (\ref{ec16}) above. We can calculate the third element of the tetrad, $\Lambda_2^\mu$, as follows
\begin{equation}
\begin{split}
(\Lambda_2^\mu)_{nn}&=J^\mu-\frac{J^\nu U_\nu}{U^\nu U_\nu}U^\mu-\frac{J^\nu A_\nu}{ A^\nu A_\nu}A^\mu\\
&=J^\mu-(U_\nu J^\nu) U^\mu-\frac{A_\nu J^\nu}{(-\alpha^2)}A^\mu~.
\end{split}
\end{equation}
Taking the derivative of \(U_\mu U^\mu=1\) gives
$U_\mu A^\mu=0~.$ Differentiating this relation once more gives $U_\mu J^\mu=-A_\mu A^\mu=\alpha^2$. Similarly, differentiating \(A_\mu A^\mu=-\alpha^2\) gives
$A_\mu J^\mu=-\alpha\dot{\alpha}~.$
Consequently,
\begin{equation}
(\Lambda_2^\mu)_{nn}
=
J^\mu-\alpha^2U^\mu-\frac{\dot{\alpha}}{\alpha}A^\mu~.
\end{equation}
Defining
\[
W^\mu
\equiv
J^\mu-\alpha^2U^\mu-\frac{\dot{\alpha}}{\alpha}A^\mu,
\]
its norm is
\begin{equation}
W_\mu W^\mu
=
J_\mu J^\mu-\alpha^4+\dot{\alpha}^2~.
\end{equation}
Therefore, imposing the normalization condition
$\Lambda_{2\mu}\Lambda_2^\mu=-1$, one obtains
\begin{equation}
\Lambda_2^\mu=
\frac{
J^\mu-\alpha^2U^\mu-\frac{\dot{\alpha}}{\alpha}A^\mu
}{
\sqrt{\alpha^4-\dot{\alpha}^2-J_\nu J^\nu}
}~.
\end{equation}
We can now make the substitutions in the FS equations. For $a=0$ we have $\dot{\Lambda}_0^\mu=K_0^b\Lambda_b^\mu$, it is obtained that $\kappa(s)=\alpha(s)$ and $\dot{U}^\mu=A^\mu$, this is the same result as in Section \ref{sec3}, the curvature is the proper acceleration. Similarly, for $a=1$ we have $\dot{\Lambda}_1^\mu=K_1^b \Lambda_b^\mu$ which is $\dot{\Lambda}_1^\mu=\kappa U^\mu+\tau \Lambda_2^\mu$, and considering Eq.~(\ref{ec17}), we write it as
\begin{equation}
\frac{J^\mu}{\alpha}-\frac{\dot{\alpha}}{\alpha^2}A^\mu=\alpha U^\mu+\tau\left(\frac{J^\mu-\alpha^2 U^\mu-\frac{\dot{\alpha}}{\alpha}A^\mu}{\sqrt{\alpha^4-\dot{\alpha}^2-J_\nu J^\nu}}\right)~,
\end{equation}
so that $\sqrt{\alpha^4-\dot{\alpha}^2-J_\nu J^\nu}=\alpha\tau$, which by taking $\alpha=\kappa$ becomes
\begin{equation}\label{modjerktor}
J_\nu J^\nu=\kappa^4-\dot{\kappa}^2-\kappa^2\tau^2~.
\end{equation}
For the index $a=2$, we need the expression for $\dot{\Lambda}_2^\mu$, which is found to be
\begin{equation}\label{v2mu}
\dot{\Lambda}_2^\mu=\frac{1}{\alpha\tau}S^\mu-\frac{2\dot{\alpha}\tau+\alpha\dot{\tau}}{\alpha^2\tau^2}J^\mu+\frac{2\dot{\alpha}^2\tau+\alpha\dot{\alpha}\dot{\tau}-\alpha\ddot{\alpha}\tau-\alpha^4\tau}{\alpha^3\tau^2}A^\mu+\frac{\alpha\dot{\tau}-\dot{\alpha}\tau}{\tau^2}U^\mu~,
\end{equation}
so, for $\dot{\Lambda}_2^\mu=K_2^1\Lambda_1^\mu=-\tau \Lambda_1^\mu$, and considering $\alpha(s)=\kappa(s)$, one can write
\begin{equation}\label{frenserrkt}
\begin{split}
&\dot{\Lambda}_2^\mu+\frac{\tau}{\alpha}A^\mu=0,\\
&S^\mu-\frac{2\dot{\kappa}\tau+\kappa\dot{\tau}}{\kappa\tau}J^\mu+\frac{2\dot{\kappa}^2\tau+\kappa\dot{\kappa}\dot{\tau}-\kappa\ddot{\kappa}\tau-\kappa^4\tau+\kappa^2\tau^3}{\kappa^2\tau}A^\mu+\frac{\kappa}{\tau}(\kappa\dot{\tau}-\dot{\kappa}\tau)U^\mu =0~,
\end{split}
\end{equation}
which can be written as
\begin{equation}\label{eq.33}
S^\mu-(2\sigma_\kappa+\sigma_\tau)J^\mu+(\sigma_\kappa^2+\sigma_\kappa\sigma_\tau-\dot{\sigma}_\kappa-(\kappa^2-\tau^2))A^\mu+\kappa^2(\sigma_\tau-\sigma_\kappa)U^\mu=0,
\end{equation}
where $\sigma_\kappa=\dot{\kappa}/\kappa$ and $\sigma_\tau=\dot{\tau}/\tau$, and we used $\dot{\sigma}_\kappa=\ddot{\kappa}/\kappa-\sigma^2_\kappa$.\\

Eq.~(\ref{frenserrkt}) is a set of three third-order differential equations that can be separated into one time and two spatial components. We directly consider restrictions on the jerk, since we can parameterize the equations, once developed, in terms of the FS frame.

\subsection{Null vector case: $||J||^2=0$} 

We now move to analyze a particular case of the FS equations with non-constant proper acceleration. From Eq.~(\ref{modjerktor}) we have
\begin{equation}\label{eq36}
\kappa^4-\dot{\kappa}^2-\kappa^2\tau^2=0~,
\end{equation}
which can be written as
\begin{equation}\label{ec37}
\frac{d\kappa}{ds}=\kappa\sqrt{\kappa^2-\tau^2}~.
\end{equation}
We consider three possibilities depending on the value of $\tau$.
\begin{itemize}
\item $\tau=0$, corresponds to the case with the single FS parameter, $\kappa$, from the previous section.
\item If $\tau$ is a nonzero constant, then one can write
\begin{equation}
\int\frac{d\kappa}{\kappa\sqrt{\kappa^2-\tau^2}}=\int ds+\bar{c}_1~,
\end{equation}
which has a solution of the form
\begin{equation}
\frac{1}{\tau}{\rm arctan \ }\left(\frac{\sqrt{\alpha^2-\tau^2}}{\tau}\right)=s+\bar{c}_1~,
\end{equation}
where $\bar{c}_1$ is an arbitrary integration constant. 
Solving for $\kappa$, one obtains 
\begin{equation}\label{tausec}
\kappa(s)=\tau\sec[\tau(s+\bar{c}_1)]~.
\end{equation}
An initial condition that provides $\bar{c}_1=0$ implies $\kappa(s=0)=\kappa_0=\tau$. We consider the initial condition $\kappa(s=0)=\kappa_0=\tau\sec(\bar{c}_1\tau)$, which gives
\begin{equation}\label{alfaj0}
\bar{c}_1=\frac{1}{\tau}{\rm arcsec \ }\left(\frac{\kappa_0}{\tau}\right)~.
\end{equation}
However, this initial condition implies $\kappa_0\neq 0$, since otherwise $\bar{c}_1$ diverges.  

On the other hand, these considerations also modify the FS Eq.~(\ref{frenserrkt}). 
Taking the derivative of Eq.~(\ref{ec37}), one obtains
\begin{equation}
\ddot{\kappa}=2\kappa^3-\kappa\tau^2~,
\end{equation}
so we can write the FS Eq.~(\ref{frenserrkt}) in the form
\begin{equation}
S^\mu-2\sqrt{\kappa^2-\tau^2}J^\mu-\kappa^2A^\mu-\kappa^2\sqrt{\kappa^2-\tau^2}U^\mu=0~.
\end{equation}
For this particular case, since $||J||^2=0$, it is easy to compute the snap modulus as
\begin{equation}
    ||S||^2=-\kappa^4\tau^2.
\end{equation}

\item The last possibility we evaluate is that $\tau(s)$ is a function of $s$ proportional to $\alpha(s)=\kappa(s)$
\begin{equation}
\tau=p\kappa~,
\end{equation}
where $p$ is a positive real constant. In this case, one has
\begin{equation}
\kappa(s)=-\frac{1}{\sqrt{1-p^2}s+c_2},
\end{equation}
where $c_2$ is an integration constant determined from the initial conditions. The previous equation also limits $0<p<1$ for $\kappa(s)$ to be real. An initial condition of the form $\kappa(s=0)=\kappa_0$ gives $c_2=-1/\kappa_0$, so we have
\begin{equation}
\kappa(s)=\frac{\kappa_0}{1-\kappa_0\sqrt{1-p^2}s}.
\end{equation}
This particular initial condition makes $\kappa$ have a range of positive values if $s<1/(\kappa_0\sqrt{1-p^2})$. \\

For this particular case, we have $\dot{\tau}=p\dot{\kappa}$, and since $||J||^2=\kappa^4-\dot{\kappa}^2-\kappa^2\tau^2=0$, we have
\begin{equation}
\dot{\kappa}=\sqrt{1-p^2}\kappa^2.
\end{equation}
The previous equation can be differentiated with respect to $s$ to obtain
\begin{equation}
\ddot{\kappa}=2(1-p^2)\kappa^3.
\end{equation}
Taking into account $\tau$, $\dot{\kappa}$ and $\ddot{\kappa}$ in Eq.~(\ref{frenserrkt}) it is found that
\begin{equation}
S^\mu-3\sqrt{1-p^2}\kappa J^\mu=0,
\end{equation}
and since $||J||^2=0$ is very easy to compute the snap modulus as
\begin{equation}
    ||S||^2=0.
\end{equation}

One can observe that every case for $J$ and $\tau$ changes $\kappa(s)$ considerably, where, in some cases, it cannot be solved analytically. These considerations for $J$ and $\tau$ also change the form of the Frenet-Serret equation, simplifying it.

\end{itemize}

\begin{figure}[h]
\includegraphics[scale=0.5]{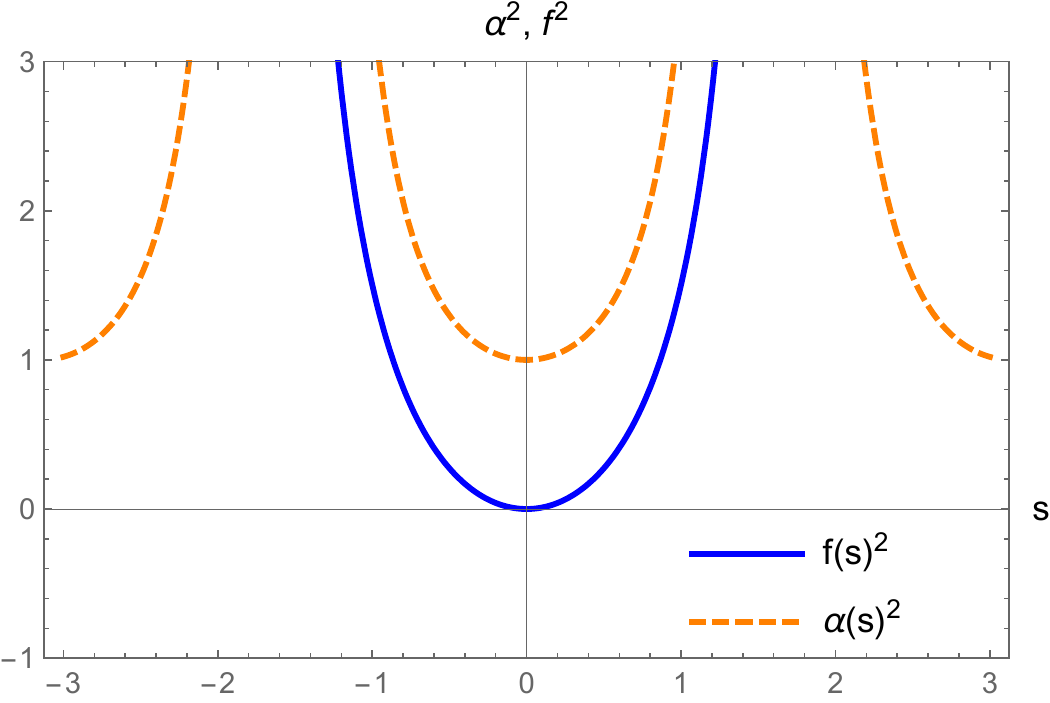}
\includegraphics[scale=0.5]{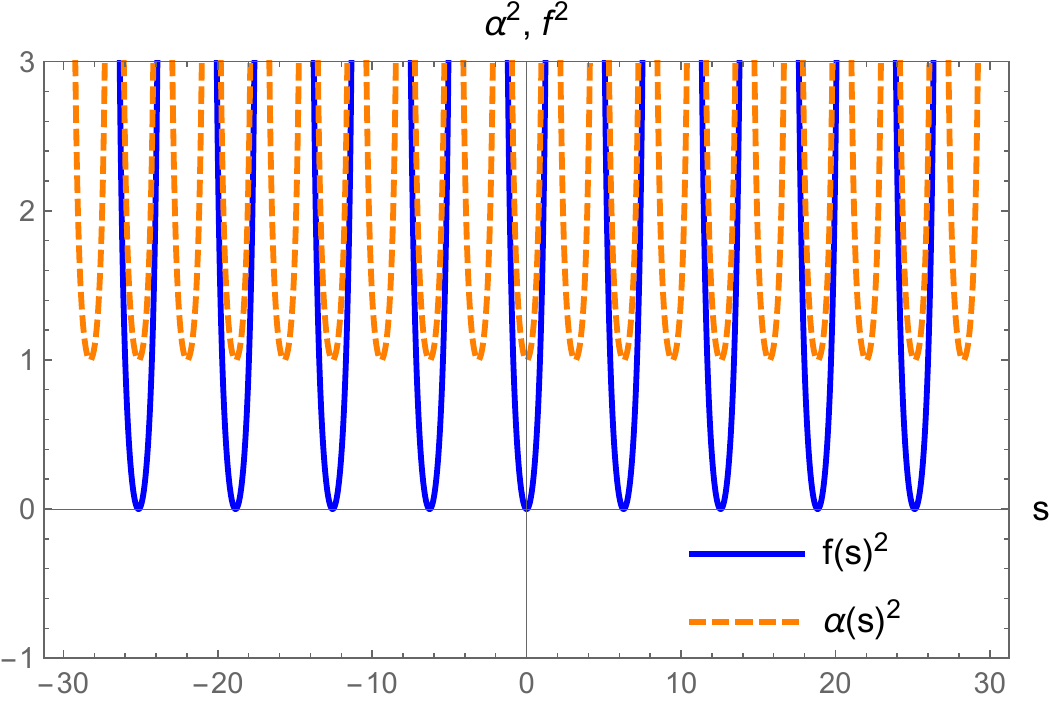}
\caption{Plots of $\alpha^2$ and $f^2$, the square of acceleration and rapidity, respectively, with $s\in(-3,3)$ and $s\in(-30,30)$, and $\tau=1$ and $\bar{c}_1 = 0$, given by Eq.~(\ref{tausec}) and its integration for the rapidity. The dynamics have periodicity in the proper time of both proper acceleration and rapidity.}
\label{sn}
\end{figure}

\subsection{Case: $||J||^2$ non-zero constant}\label{ssec4.2} 

We can express Eq.~(\ref{modjerktor}) as
\begin{equation}
\label{intj2c}
\int\frac{d\kappa}{\sqrt{\kappa^4-\tau^2\kappa^2-||J||^2}}=\int ds+c_1~,
\end{equation}
where again, the results of the evaluation of this integral depend on the values taken by the torsion parameter $\tau$.
\begin{itemize}
\item If $\tau=0$ we return to the case where we only had $\kappa$, which was an elliptic sn function.
\item If $\tau$ is a non-zero constant, we obtain an elliptic function of the first class
\begin{equation}
\frac{(\sqrt{4||J||^2+\tau^4}-\tau^2)^{1/2}}{\sqrt{2}||J||}{\rm F}~\left(i~{\rm arcsinh}~\left(\kappa\sqrt{\frac{2}{\sqrt{4||J||^2+\tau^4}-\tau^2}}\right),\frac{\tau^2-\sqrt{4||J||^2+\tau^4}}{\tau^2+\sqrt{4||J||^2+\tau^4}}\right)=s+c_1~,
\end{equation}
Similarly to previous cases, one can find $\kappa$ in terms of $s$ as
\begin{equation}\label{Eq 53}
\kappa(s)=i\sqrt{\frac{\sqrt{4||J||^2+\tau^4}-\tau^2}{2}}\text{sn}\left(\frac{s \sqrt{\frac{\sqrt{4 ||J||^2+\tau ^4}-\tau ^2}{||J||^2}} \left(\sqrt{4 ||J||^2+\tau ^4}+\tau ^2\right)}{2 \sqrt{2}}, \frac{\tau
   ^2-\sqrt{4 ||J||^2+\tau ^4}}{\sqrt{4 ||J||^2+\tau ^4}+\tau ^2}\right)~,
\end{equation}
where $\kappa(s)$ always has purely imaginary values. For simplicity, $c_1=0$ has been considered.\\

On the other hand, the FS equation is also modified. Since $||J||^2$ is a constant, $\alpha$ is found to be
\begin{equation}
\dot{\kappa}=\sqrt{\kappa^4-\kappa^2\tau^2-||J||^2},
\end{equation}
and deriving the previous equation, one has
\begin{equation}
\ddot{\kappa}=2\kappa^3-\kappa\tau^2.
\end{equation}
Considering $\dot{\kappa}$, $\dot{\tau}$ and $\ddot{\kappa}$ in Eq.~(\ref{frenserrkt})
\begin{equation}
\kappa^2 S^\mu-2\kappa\sqrt{\kappa^4-\kappa^2\tau^2-||J||^2}J^\mu-(2||J||^2+\kappa^4)A^\mu-\kappa^3\sqrt{\kappa^4-\kappa^2\tau^2-||J||^2}U^\mu=0.
\end{equation}

\begin{figure}[h]
\includegraphics[scale=0.7]{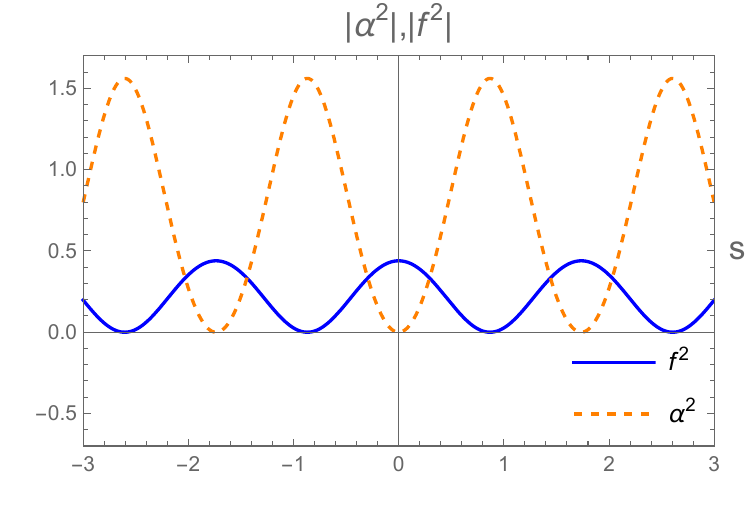}
\includegraphics[scale=0.7]{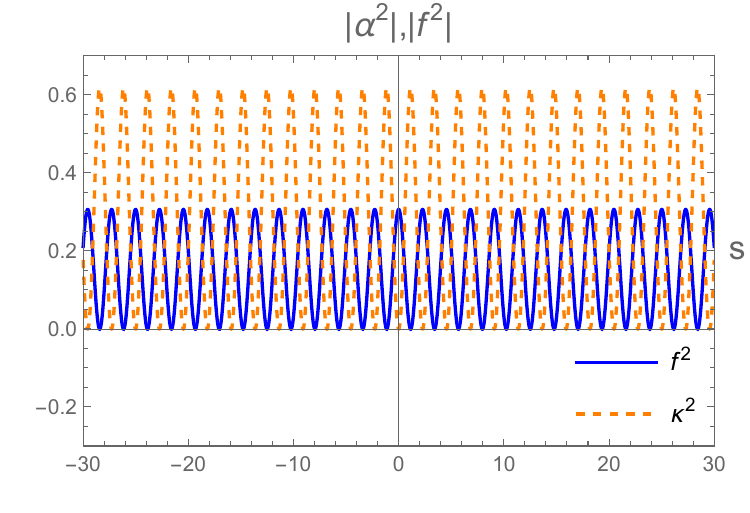}
\caption{Plots of $|\alpha|^2$ and $|f|^2$, the modulus of acceleration and rapidity, respectively, with $s\in(-3,3)$, $\tau=1$ and $||J|| = 1$, given by Eq.~(\ref{Eq 53}) and its integration for the rapidity. The dynamics have periodicity in the proper time of both proper acceleration and rapidity, and the behavior is very similar to the case $||J||^2=0$, but since we now have complex values the modulus was plotted.}
\label{sn2}
\end{figure}

\item The last case of $||J||^2=$ constant $\neq 0$ is when $\tau$ is a non-constant parameter, but taken proportional to $\kappa$,
\begin{equation}
    \tau(s)=p\kappa(s),
\end{equation}
where $p$ is a positive real constant. Then, Eq.~(\ref{intj2c}) can be written as
\begin{equation}\label{eq57}
\int\frac{d\kappa}{\sqrt{(1-p^2)\kappa^4-||J||^2}}=s+c_1,
\end{equation}
where $c_1$ is an integration constant determined by initial conditions. The solution is given by an elliptic integral of the first kind ${\rm F}$,
\begin{equation}\label{eq58}
-\frac{i}{\sqrt{||J||}(1-p^2)^{1/4}}
{\rm F}\left(
\arcsin\left[\frac{(1-p^2)^{1/4}}{\sqrt{||J||}}\kappa\right],
-1
\right)=s+c_1~.
\end{equation}
The inverse function, $\kappa$ in terms of $s$, is given analytically by the Jacobi sine function
\begin{equation}\label{Eq. 64}
\kappa(s)=
\sqrt{\frac{||J||}{\sqrt{1-p^2}}}\,
{\rm sn}\left(i\sqrt{||J||\sqrt{1-p^2}}(s+c_1),-1\right),
\end{equation}
where $\kappa(s)$ is always purely imaginary for this analytic branch.\\

Since
\begin{equation}
||J||^2=\kappa^4-\dot{\kappa}^2-\kappa^2\tau^2
    =(1-p^2)\kappa^4-\dot{\kappa}^2
\end{equation}
is a constant, one can obtain $\dot{\kappa}$ as
\begin{equation}
\dot{\kappa}=\sqrt{(1-p^2)\kappa^4-||J||^2},
\end{equation}
and consequently
\begin{equation}
\ddot{\kappa}=2(1-p^2)\kappa^3.
\end{equation}
Since $\tau=p\kappa$, one has $\dot{\tau}=p\dot{\kappa}$, and therefore the $U^\mu$ term in Eq.~(\ref{frenserrkt}) vanishes. This modifies FS Eq.~(\ref{frenserrkt}) as
\begin{equation}
S^\mu-\frac{3}{\kappa}\sqrt{(1-p^2)\kappa^4-||J||^2}J^\mu-\frac{3||J||^2}{\kappa^2}A^\mu=0.
\end{equation}

\end{itemize}

We have not yet used a parameterization, but we can consider one in terms of a hyperbolic one $U^\mu\to (\cosh(f(s)),\vec{n}(s)\sinh(f(s)))$, where $\vec{n}$ is a two-vector with unitary magnitude \cite{ponspalol2019}. The time component in Eq.~(\ref{frenserrkt}) can be written as
\begin{equation}
\begin{split}
&\sinh(f)\left[
\dddot{f}+(\dot{f})^3
-\frac{2\dot{\kappa}\tau+\kappa\dot{\tau}}{\kappa\tau}\ddot{f}
+\frac{
2\dot{\kappa}^2\tau
+\kappa\dot{\kappa}\dot{\tau}
-\kappa\ddot{\kappa}\tau
-\kappa^4\tau
+\kappa^2\tau^3
}{
\kappa^2\tau
}\dot{f}
\right]\\
&+
\cosh(f)\left[
3\dot{f}\ddot{f}
-\frac{2\dot{\kappa}\tau+\kappa\dot{\tau}}{\kappa\tau}\dot{f}^2
+\frac{\kappa}{\tau}(\kappa\dot{\tau}-\dot{\kappa}\tau)
\right]
=0.
\end{split}
\end{equation}
A particularly simple reduction occurs if the logarithmic derivatives of the
curvature and torsion are opposite,
\begin{equation}\label{condition}
    \frac{\dot{\tau}}{\tau}=-\frac{\dot{\kappa}}{\kappa}.
\end{equation}
This condition implies
\begin{equation}
    \kappa(s)\tau(s)=C,
\end{equation}
where \(C\) is a constant. Equivalently, after choosing units such that
\(C=1\), one has
\begin{equation}
    \kappa(s)=\frac{1}{\tau(s)}.
\end{equation}
Under this reciprocal curvature-torsion condition, Eq.~(\ref{eq.33})
simplifies to
\begin{equation}
    S^\mu-\sigma_\kappa J^\mu
    -\left(\dot{\sigma}_\kappa+\frac{\kappa^4-C^2}{\kappa^2}\right)A^\mu
    -2\sigma_\kappa\kappa^2 U^\mu=0.
\end{equation}

\begin{figure}[h]
\includegraphics[scale=0.7]{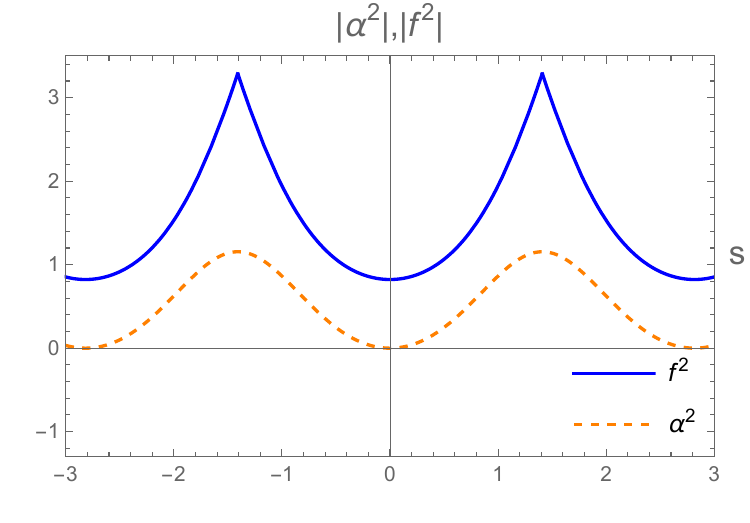}
\includegraphics[scale=0.7]{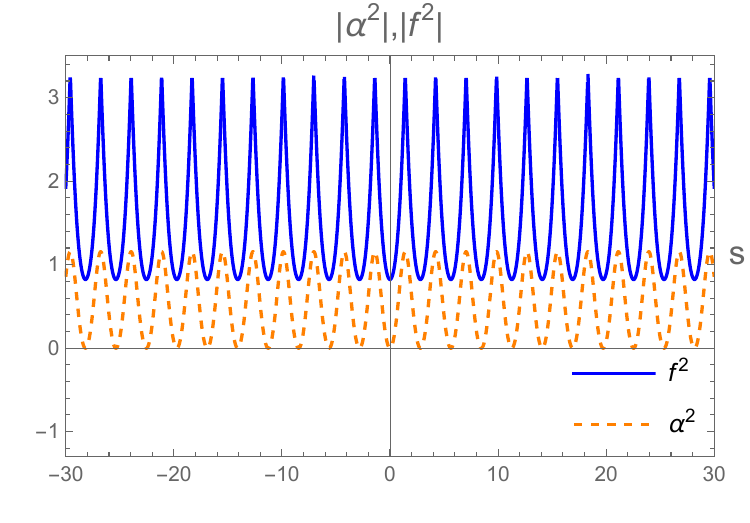}
\caption{Plots of $|\alpha|^2$ and $|f|^2$, the modulus of acceleration and rapidity, respectively, with $s\in(-3,3)$, $p=0.5$ and $||J|| = 1$, given by Eq.~(\ref{Eq. 64}) and its integration for the rapidity. The dynamics exhibit periodicity in proper time for both proper acceleration and rapidity, and the behavior is very similar to previous cases, with the modulus plotted instead of the square values.}
\label{sn2}
\end{figure}

\section{Conclusions}\label{sec5}

We have studied relativistic motion with non-constant proper acceleration by using the generalized Frenet-Serret frame in Minkowski spacetime. The analysis was developed step by step, beginning with the case in which the only nonzero Frenet-Serret invariant is the curvature, and then extending the discussion to include torsion. In both cases, the Frenet-Serret construction gives a direct geometrical interpretation of the usual kinematic quantities: the curvature is identified with the proper acceleration, while the jerk and higher derivatives are constrained by the curvature, torsion, and their proper-time derivatives.

A central point of the analysis is that, in relativistic motion, the behavior of the four-jerk is not consistent with the Newtonian intuition based solely on the derivative of the magnitude of the acceleration. In particular, even when the proper acceleration is constant, the four-jerk need not vanish. For uniform proper acceleration, one has $\dot{\alpha}=0$, but the four-jerk is still related to the four-velocity, and its invariant satisfies $J^2=\alpha^4$. Therefore, a nonzero constant proper acceleration corresponds to a nonzero constant jerk invariant. Conversely, for constant proper acceleration, the condition of vanishing jerk invariant forces the proper acceleration itself to vanish. This emphasizes that the invariant $J^2$ is a genuinely relativistic quantity, sensitive not only to changes in the magnitude of the acceleration but also to the hyperbolic rotation of the four-velocity and four-acceleration in spacetime.

For the curvature-only case, the Frenet-Serret equations lead to the simple relation
$J^2=\kappa^4-\dot{\kappa}^2$. This relation allows one either to determine the jerk invariant from a prescribed proper acceleration, or conversely to obtain a differential equation for the curvature once the jerk invariant is specified. The case $J^2=0$ yields a real, non-uniform proper acceleration of the form $\kappa(s)\propto (s+c_1)^{-1}$, leading to trajectories that differ from the usual Rindler hyperbola yet remain expressible analytically. For nonzero constant $J^2$, the curvature is naturally expressed in terms of Jacobi elliptic functions, and the corresponding rapidity is obtained by integrating the curvature. In this case the analytic branch considered here is complex, so the usual purely hyperbolic parametrization of the velocity is no longer strictly real.

Including torsion enriches the structure of the Frenet-Serret system. The jerk invariant becomes
\[
J^2=\kappa^4-\dot{\kappa}^2-\kappa^2\tau^2,
\]
showing explicitly that torsion contributes to the invariant balance between curvature, the rate of change of curvature, and jerk. Thus, in the torsionful case, the jerk is no longer determined solely by the proper acceleration. Instead, it also measures how the worldline departs from motion confined to the two-dimensional plane generated by the four-velocity and four-acceleration. In this sense, torsion encodes additional spacetime geometry of the trajectory, and its presence changes both the allowed forms of $\kappa(s)$ and the simplified Frenet-Serret equations obtained under restrictions on $J^2$.

Several explicit cases were considered. For $J^2=0$ with constant torsion, the curvature is given by a secant function, while for torsion proportional to curvature, $\tau=p\kappa$, the curvature has a rational dependence on proper time. For a nonzero constant $J^2$, both the constant-torsion and proportional-torsion cases yield elliptic-function expressions for the curvature. These examples show that simple assumptions about the invariant jerk and torsion can yield nontrivial yet analytically tractable classes of relativistic trajectories.

The rapidity function $f(s)$ can be written as the integral of the proper acceleration in the curvature-only case, and this remains a useful guide for interpreting the motion. However, once torsion is present, a fixed-direction hyperbolic parametrization of the form $U^\mu=(\cosh f,\hat n\sinh f)$ is generally too restrictive unless additional conditions are imposed. In the torsionful case, the spatial direction of the motion must, in general, evolve, reflecting that torsion describes the rotation of the Frenet-Serret frame beyond the plane of the velocity and acceleration.

Finally, we found that the Frenet-Serret equations are substantially modified by the assumptions imposed on $J^2$ and $\tau$. In particular, the term accompanying the four-jerk often has a structure closely related to that of the term accompanying the four-velocity, differing only by powers of the proper acceleration. This is consistent with the curvature-only relation for uniform acceleration, where the four-jerk is proportional to the four-velocity, $J^\mu=\alpha^2U^\mu$. A vanishing jerk invariant corresponds to a null four-jerk, for which the temporal and spatial contributions to the invariant cancel. This provides a useful analogy with other null relativistic quantities, such as those appearing in the description of electromagnetic radiation.

\section{Acknowledgements} Funding comes partly from the FY2024-SGP-1-STMM Faculty Development Competitive Research Grant (FDCRGP) no.201223FD8824 and SSH20224004 at Nazarbayev University in Qazaqstan.  Appreciation is given to the ROC (Taiwan) Ministry of Science and Technology (MOST), Grant no. 112-2112-M-002-013, the National Center for Theoretical Sciences (NCTS), and the Leung Center for Cosmology and Particle Astrophysics (LeCosPA) of National Taiwan University.
\bibliography{main} 
\end{document}